\documentclass[12pt,preprint]{aastex}
\usepackage{psfig}

\newcommand{\citepeg}[1]{\citep[{e.g.,}][]{#1}}

\def\lsim{\hbox{ \rlap{\raise 0.425ex\hbox{$<$}}\lower 0.65ex\hbox{$\sim$} }}
\def\gsim{\hbox{ \rlap{\raise 0.425ex\hbox{$>$}}\lower 0.65ex\hbox{$\sim$} }}
\def\arcmin{\hbox{$^\prime$}}
\def\arcsec{\hbox{$^{\prime\prime}$}}

\def\f(h{\hbox{$~\!\!^{\rm h}$}}

\def\ale{\mathrel{\hbox{\rlap{\hbox{\lower4pt\hbox{$\sim$}}}\hbox{$<$}}}}
\def\age{\mathrel{\hbox{\rlap{\hbox{\lower4pt\hbox{$\sim$}}}\hbox{$>$}}}}

\shorttitle{Optical Precursor Limits on GRBs}
\begin{document}

\title{Optical Limits on Precursor Emission from Gamma-Ray Bursts with
Known Redshift}

\author{C.~Blake}

\affil{Department of Astrophysical Sciences, Princeton University, Princeton, 
NJ 08544, USA cblake@astro.princeton.edu}

\author{J.~S.~Bloom}

\affil{Harvard-Smithsonian Center for Astrophysics,  
60 Garden Street, Cambridge, MA 02138, USA jbloom@cfa.harvard.edu}

\affil{Harvard Society of Fellows, 78 Mount Auburn Street, Cambridge, MA 02138 USA}

\begin{abstract}
Making use of virtual observatory data, we present the first
comprehensive sample of optical observations conducted before the
explosion times of all gamma-ray bursts (GRBs) with known
redshifts. In total, the fields of 11 such GRBs were observed by the
Near-Earth Asteroid Tracking (NEAT) program from years to hours before
the bursts. Though the typical limiting magnitudes from these
observations are $R \approx 20$\,mag, we find no evidence for a
significant detection of a precursor. The deepest non-detection of
precursor emission is from GRB\,030329, reaching down to an absolute
$B$-band magnitude of $M_B \approx -18$\,mag from 6--1500 days
(restframe) before the burst. This is of comparable brightness to
supernovae, which, in some scenarios for GRB progenitors, are
predicted to pre-date a GRB on similar timescales.  Since sources
cannot be localized to better than $\sim$500 milliarcsecond (3
$\sigma$) with current large-area surveys, unrelated supernovae or AGN
activity in GRB hosts could be mistaken for genuine precursor
emission. This possibility motivates the need for not only deep
wide-field imaging, but imaging at high spatial resolution.
\end{abstract}

\keywords{cosmology: observations --- gamma rays: bursts}

\section{Introduction}

Constraints on the character and occurrence of precursor emission from
gamma-ray bursts (GRBs) offer insight into the nature of the
explosions. On timescales of hours to seconds before a burst,
precursor emission could occur from the central engines themselves,
by, for example, blackbody emission from the early fireball
\citep{lp98b,lp99,lu00,dai02} or a pre-fireball created in the merger
of neutron stars \citep{rrr02}. If GRBs were to occur from stellar
disruption around a massive black hole \citep{car92}, then thermal
precursors ($T \sim 50$ eV) which decay as power-laws \citepeg{lnm02}
might be expected if other stars are disrupted by the same black
hole.

If GRBs arise from the death of massive stars, the progenitor star (or
its companion, if the progenitor is in a binary system) could produce
precursor emission in the form of novae or supernovae (SNe). A precursor SN
is a generic prediction of the so-called ``supranova'' scenario
\citep{vs98,vs99}. In the supranova scenario, a massive stellar
progenitor explodes to produce an SN, leaving behind a
supermassive spinning neutron star. The neutron star eventually
collapses to form a black hole and a GRB. \citet{vs98} posited the
supranova picture as a means to pre-evacuate the space around the GRB
explosion and avoid baryonic contamination to the relativistic
blastwave. Aside from this theoretical consideration, the supranova
scenario is attractive for two observational reasons. First, the
sequence of events in this scenario is a natural means to explain
transient X-ray features attributed to dense sub-relativistic matter
streaming away at distance of $10^{14}$--10$^{17}$\,cm from the
explosion site
\citepeg{pggs+00,afv+00,rwo+02}. Second, the gas behind the SN shock
can be thermalized by the time the GRB fireball sweeps through,
leading to an apparent constant density medium in afterglow modeling
\citep{kg02,igp03}. This would then explain why GRBs from massive
stars would typically not show evidence for a wind-blown medium
surrounding most bursts \citepeg{cl99,pk01b}. While the significance
of some of the X-ray features has been questioned
\citepeg{rs03} and winds are generally not ruled out statistically 
through afterglow modeling \citep{pk01b}, the possibility of a
precursor SN on timescales of days \citep{dkrw02} to years is a prime
motivation to search for precursor emission from GRBs.

As summarized in \cite{hpb94}, archival photographic plate searches
\citepeg{sch83} for optical precursors (in the context of recurrence 
of counterpart emission) revealed only a few genuine astrophysical
sources, with no definitive establishment of a physically-associated
precursor to any GRB. Using pre-images of the fields of seven {\em
Burst and Transient Source Experiment} (BATSE) GRBs from the Explosive
Transient Camera, \citet{kvr96} found no precursors to $V \approx 6$\,mag. 
\citet{gwh+95} presented eight upper limits to precursor
emission, at $V = 11-13$\,mag, within 12 hrs preceding BATSE
bursts. Had a genuine transient been found before a GRB, the large GRB
error boxes (typically much larger than tens of sq.~arcminutes) would
have required an association based upon probabilistic arguments. Since
the chance for unrelated transients to occur in GRB error boxes is
non-negligible, the large BATSE error boxes all but prevented an
unambiguous association with the GRBs themselves.

Since 1997, over forty GRBs have been localized to the sub-arcsecond
level by way of transient afterglow emission. This, coupled with the
growing number of large-area survey projects ($\age \pi$ steradian),
dramatically increases the likelihood of deep pre-imaging of GRB
positions. Pre-imaging from near-Earth asteroid surveys allowed
\citet{gyof+02} to show that the transient optical source SDSS
J124602.54+011318.8 was not an orphan afterglow
\citep{vblw+02}, but instead was a rare class of AGN.  Recently,
\citet{heyl03} suggested that with an increased frequency of GRB
localizations from {\it Swift} \citep{geh00} and very deep large-field
optical imaging surveys such as the Sloan Digital Sky Survey
(SDSS; \citealt{yaa+00}) and PanSTARRS \citep{kai02}, constraints
could routinely be placed on SN (or, more generally, SN-like) precursors to 
GRBs.

Other than progenitor scenarios where a SN precedes a GRB, we stress
that few models exist that make specific predictions (e.g., of light
curves, spectra, start times relative the GRB) for precursor
emission. To this end, we have undertaken an exploratory study of
public pre-imaging to search for any optical precursor transients at GRB
positions.  Here, we present pre-imaging observations of 11 GRBs. By
focusing on those bursts with known redshift we can translate the
derived upper limits into absolute magnitudes in co-moving time.  In
most bursts, the upper limit to optical precursor emission is
$\sim$3--5 magnitudes brighter than the brightest supernovae known, but
comparable to afterglow brightnesses. These derived upper limits are
consistent with the known (fainter) brightnesses of the respective GRB
host galaxies, later found at the transient positions.

\section{Observations and Analysis}

Most of the pre-imaging data presented herein were obtained by the
NEAT program \citep{prh+99}. From December 1995 to February 1999, the
NEAT camera was mounted on the 1-m GEODSS telescope in Haleakala, Maui,
Hawaii. From February 2000 onward, the camera was then mounted on the
Maui Space Surveillance Site (MSSS) 1.2-m telescope. Another NEAT
camera began operation on the Oschin 48 inch Schmidt telescope at
Palomar Observatory. Briefly, the NEAT imager on Palomar consists of
three 4080 $\times$ 4080 1.0\arcsec/pixel CCD arrays covering
approximately 1.13 $\times$ 1.13 deg$^2$. Typical exposure times are
20--60 seconds and the images are unfiltered, giving a broad spectral
response from 4000--9000 \AA. A 60 second exposure reaches $V
\sim 19-21$\,mag (3 $\sigma$), depending on sky brightness. We also 
examined the Digital Sky Survey II (DSS-II) scans of the photographic
Palomar Sky Survey \citep{lgl+98}.

For the 30 GRBs with a known redshift, we queried the public NEAT data
archive\footnotemark\footnotetext{See {\tt
http://skyview.gsfc.nasa.gov/skymorph/}; \citet{phr+98}}, retrieving
all data which predated the respective burst trigger date. Imaging data for
11 GRBs were found in the NEAT archive, all with at least one epoch of
high enough quality to produce good photometry. The amount of
available data for each GRB ranged from 3 to 80 individual 8\arcmin
$\times$ 8\arcmin\ images with exposure times between 20 sec and 150
sec. The data for each GRB were first aligned with sub-pixel accuracy
to the same astrometric grid using bi-linear interpolation. A triangle
matching algorithm was used to determine a transformation between
each frame and a common reference image by taking into account
translation, rotation, and magnification. A WCS astrometric solution
was fit to the aligned images using reference stars from the USNO-A2.0
astrometric catalog starting from the approximate center-of-field and
plate scale. The photometric zeropoint for each image was estimated in
the $R$- and $B$-band by comparison with the magnitudes given in the
USNO-A2.0. While the photometric errors on individual catalog
magnitudes is large ($\approx0.25$\,mag), a number of stars typically
larger than 100 was utilized in order to accurately derive the
photometric zeropoint for each frame. The individual images were
photometered at the known position of the OT using apertures of radius
$1.2\times$FWHM of the seeing \citep{how89} and local sky subtraction based on an
annulus of radii 15 and 30 pixels. Photometric limits were estimated
for each image by analyzing the histogram of flux in 1500 apertures of
radius $1.2\times$FWHM of the seeing placed randomly within the image. Iterative
sigma clipping was used to reject flux values contaminated by stars in
order to determine the errors due to the sky background. The standard
deviation of the Gaussian distribution of the sky background aperture
flux values was used with the previously determined photometric
zeropoint to estimate the photometric upper limits for each
image. The aperture fluxes were measured at the OT position in both the individual epoch images and the stacked images. Tables \ref{tab:photom} and \ref{tab:limits} summarize the upper limits for all the
NEAT imaging. A subset of the NEAT data on four of these bursts have
been presented previously \citep{wval02,wv02,wv03a,wvnl03}.  Our
photometric measurements of the same data agree with those reports to
within 0.5\,mag.

\begin{figure*}[hbt]
\centerline{\psfig{file=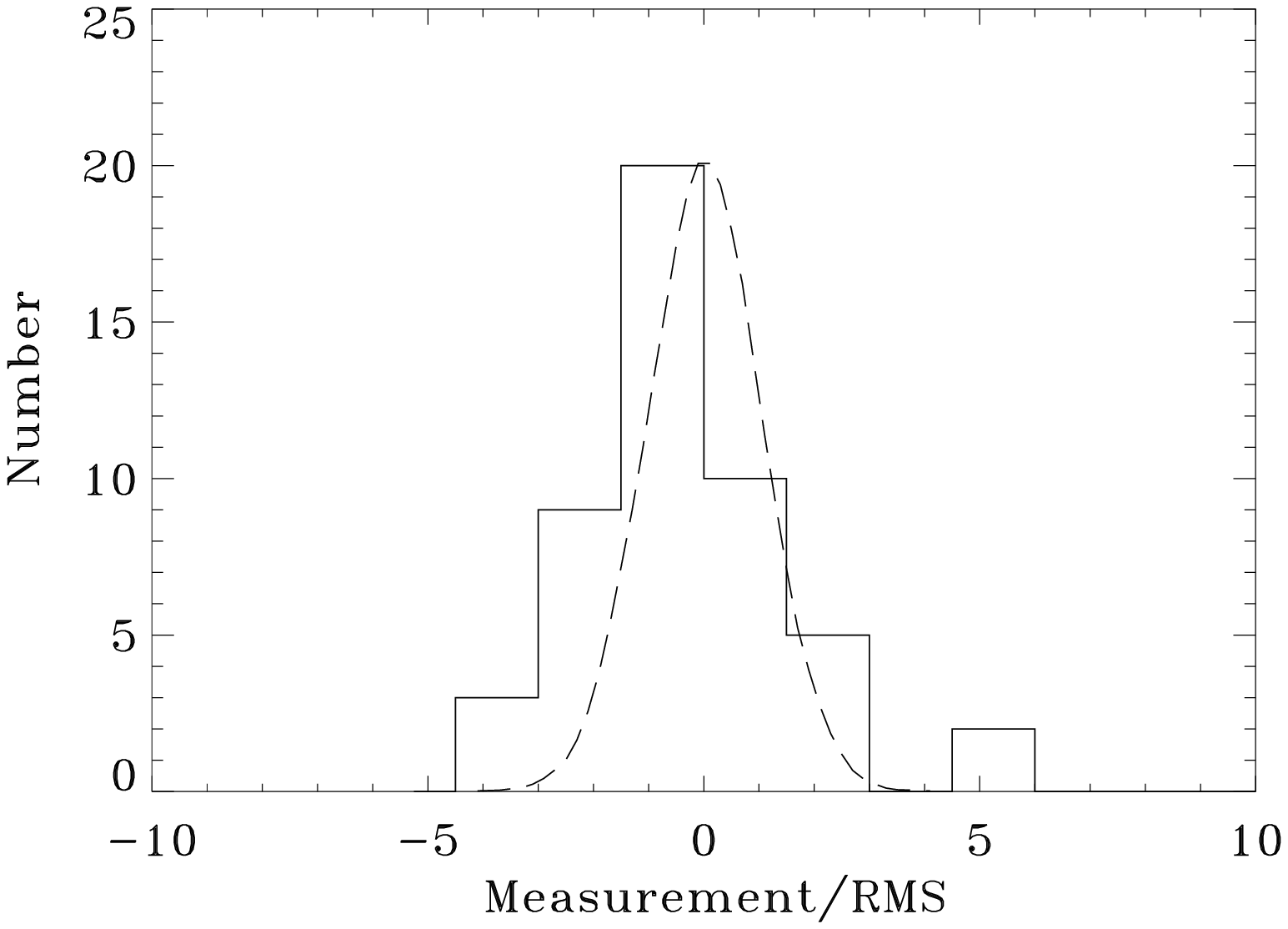,width=5.7in}}
\caption[]{Demonstrating the non-detection of precursor emission. Histogram of the rms-normalized fluxes at the OT positions for 60 single epoch observations of the 11 GRBs studied here. The dashed-curve distribution is that expected for purely Gaussian errors with perfect sky subtraction. The higher than expected number of negative flux values indicates that the sky values are sometimes overestimated. The single epoch point at $5\sigma$ is from GRB 020813 and is an artifact of poor seeing and the crowded field around the OT.}
\label{fig:hist} 
\end{figure*}

Figure \ref{fig:ims} shows the stacked preimages of the individual
fields surrounding all 11 GRBs. Before making a stacked image of a
given field, individual exposures are visually inspected for the
presence of defects near the known position of the burst and excluded
if defects or cosmic-rays are found. We photometered the 11 stacked
images at the position of the GRB and find the $R$-band upper limits
presented in Table \ref{tab:limits} and for individual epochs in Table
\ref{tab:photom}. We report $3\sigma$ upper limits on the photometry
of the individual epoch images and $2\sigma$ upper limits for the
stacked images. At this level we statistically expect to have no false
positive detections for the stacked images and at most a few
false-positives in the individual epoch photometry. The histogram of
the rms-normalized individual epoch flux measurements is shown in
Figure \ref{fig:hist}. The expected distribution from purely Gaussian errors is
also plotted. This figure indicates that our estimations of the
photometric errors are reasonably accurate, with a tendency to over
estimate local sky values, leading to a greater number of negative
flux values. A single epoch detection at the $5\sigma$ level shown in
Figure \ref{fig:hist} is attributed to an observation in very poor seeing of GRB
020813. The OT position for this GRB is in a crowded region with
several bright stars within a few NEAT pixels making accurate photometry very difficult in poor seeing. We find upper limits consistent with the known fainter brightness of the GRB host galaxies.

\section{Results}

As seen in Table \ref{tab:limits}, there are no significant detections
of precursor emission at the positions of the 11 GRBs for which a
search was conducted. To place this in the context of physically
meaningful quantities, we translate the upper limits to an approximate
absolute magnitude in the restframe $B$-band using a cosmological
$k$-correction. This $k$-correction is derived assuming an intrinsic
source spectrum of $f_\nu(\lambda) \propto \lambda^{+1}$, which
approximates the typical median $B$-$R$ colors of the photometric
calibration objects in the GRB fields (in \S \ref{sec:sne}, we vary
this assumed precursor spectrum). We use the zeropoints, zp$_{B,R}$,
and effective wavelengths, $\lambda_{B,R}$, of the photometric
bandpasses from
\citet{fsi95}. We convert the extinction corrected upper limit, $m_R$,
in the $R$-band to absolute magnitude in the $B$-band, as
$$
M_B(z) \approx m_R - DM(z) - 2.5\,\log\left(\frac{{\rm zp}_R}{{\rm zp}_B}
	\frac{\lambda_B\,(1 + z)}{\lambda_R}\right) - 2.5\,\log (1 + z),
$$
where $DM(z)$ is the distance modulus to the burst at redshift $z$,
computed assuming $H_0 = 65$ km\,s$^{-1}$\,Mpc$^{-1}$, $\Omega_m =
0.3$, and $\Omega_\Lambda$ = 0.7. The third term accounts for the
assumed spectrum of the precursor and the last term accounts for the
intrinsically smaller bandpass in the restframe. We also find
$M_B(z)$ by using the equivalent $B$-band upper limits and average the
two results. Since the NEAT spectral response is broad and the
intrinsic precursor spectrum is unconstrained, we expect the
systematic uncertainty in the subsequent $M_B$ measurements to be at
least $\approx$ 0.3\,mag, and probably greater for the higher redshift
bursts. If the precursor spectra are significantly more red (blue) than 
$f_\nu(\lambda) \propto\lambda^{+1}$, then the quoted upper limits will be 
systematically too deep (shallow), especially for bursts at high redshift.

\begin{figure*}[hbt]
\centerline{\psfig{file=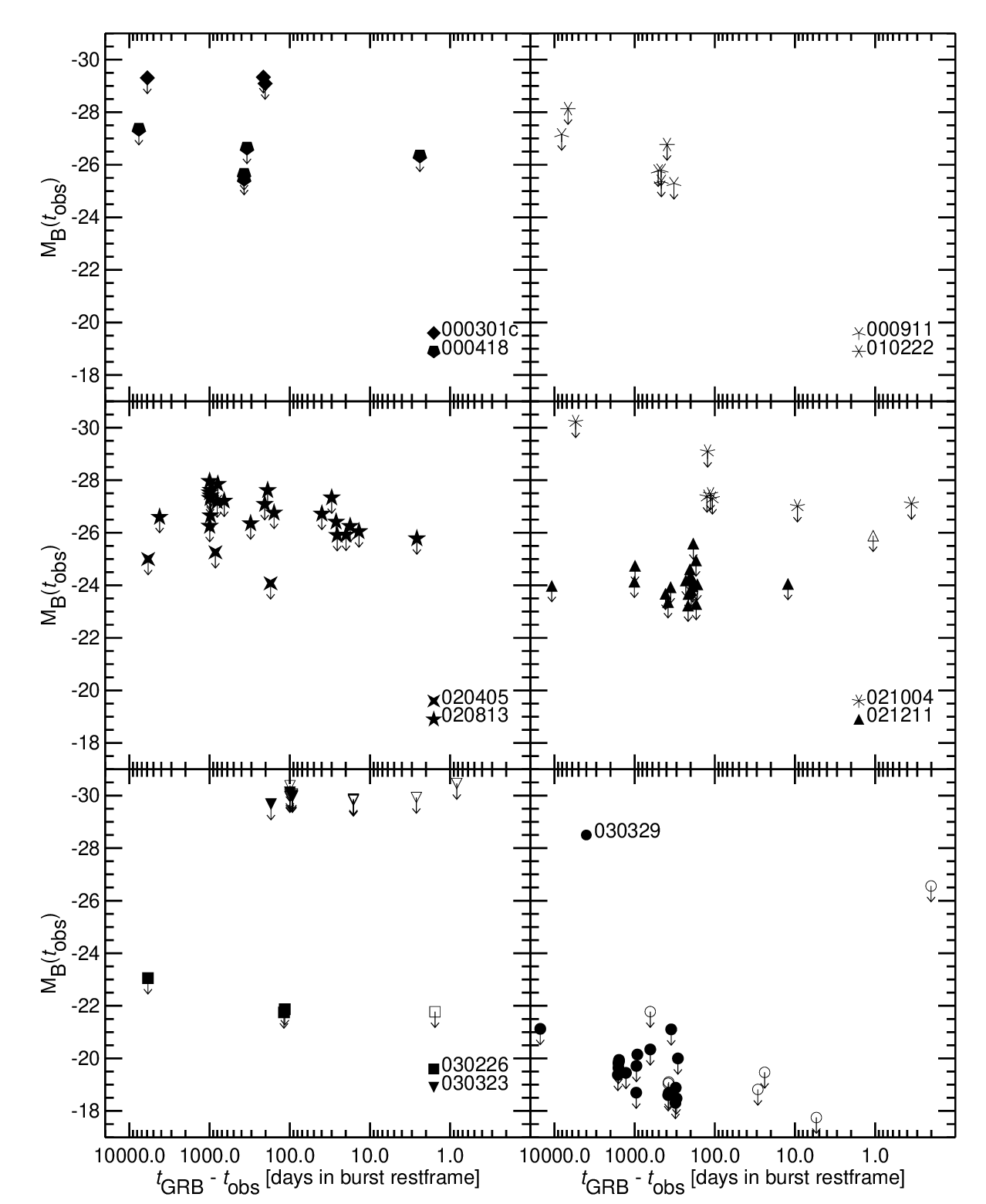,width=5.7in}}
\caption[]{Upper limits to optical precursor emission of GRBs
with known redshift. The time before the burst is given in the
restframe of the host galaxy. Absolute $B$-band magnitudes are
computed from the extinction-corrected observed optical limits as
described in the text. For all but GRB\,030329 and GRB\,030226,
long-lived optical precursors must have been fainter than $M_B \approx
-24$ to $-26$\,mag. Any SN precursor to GRB\,030329 could not have
reached brighter than $M_B({\rm peak}) \approx -19$\,mag for the 30
days prior to the explosion. Upper limits derived from the Digital Sky
Survey (DSS-II) are also shown.}
\label{fig:ups} 
\end{figure*}

Figure \ref{fig:ups} shows the aggregate absolute magnitude upper
limits from the 11 bursts with pre-imaging from NEAT and the
DSS-II. The observation time before the burst is given in the
co-moving frame of the host galaxies, i.e., the time-dilation has been
removed by dividing the observed time by (1 + $z$). Our limits are
supplemented with a few others from the literature (shown as unfilled
symbols). The early upper limit on GRB\,030329 (5.6 hours before the
burst) is from the RAPTOR experiment \citep{vbb+03,wv03}. The other
limits are 030329 \citep{wvnl03}, 030226 \citep{wv03b}, and 021211
\citep{wv02}. The deepest limit on GRB\,030329 was reported by
\citet{wvnl03} from NEAT imaging, but we were unfortunately unable to
find such images in the archive to confirm the result. In addition, a
3 minute exposure at the position of GRB\,030226 was obtained by
\citet{ctupm+03} using the BOOTES experiment \citep{ctsb+99} starting 1.5
minutes before the burst trigger. The upper limit of $R \approx 11.5$\,mag 
(not shown in Figure \ref{fig:ups}) implies any precursor would
have been $M_B \age -34.6$\,mag at the redshift of that GRB.

\subsection{Limits on Precursor Emission}
\subsubsection{GRB 030329}

At a redshift of $z=0.1685$ \citep{gpe+03}, corresponding to a
distance modulus of $DM$ = 39.70\,mag, GRB\,030329 is the nearest known
``cosmological'' GRB. Its proximity allowed several groups to study
the photometric and spectral properties of the associated supernova
with unprecedented detail
\citep{smg+03,cff+03,hsm+03,hczk03,kdw+03}. Though this firm detection
of a nearly contemporaneous SN suggests a single event destroyed the
progenitor---rendering unlikely the possibility of a similar precursor
event---it is still of interest to know what limits can be placed on
the occurrence of a similar precursor.
                                                                               
At this redshift, a bright supernova-like precursor [$M_B$(peak) =
$-20$\,mag] would have peaked at $V \sim 19.7$\,mag and could have remained
above our nominal single epoch magnitude limit of $V \sim 21$\,mag for
$\sim$2--3 weeks.  Since there are temporal gaps in the data,
particularly between 30--300 days before the GRB, it is certainly
possible that a bright precursor could have been missed by the
pre-imaging. However, three upper limits less than 30 days before the
burst exclude such a bright SN-like precursor preceding the GRB by
$\sim$45--10 days.

\subsubsection{SN-like Precursors}
\label{sec:sne}

To estimate how restrictive the entire data set is to SN-like
precursors, we created a Monte Carlo test by assuming that every GRB
in the sample had the same precursor at some random explosion date
less than 1000 days before the burst. The working hypothesis is that a
GRB with a SN-like precursor would be of type Ib/Ic and so we use SN
1998bw as a template.  Owing to the extreme deficit of flux blueward
of $\sim$3000 \AA\ of Ib/Ic SNe, we only use precursor photometry from
bursts that originated from $z < 1.2$. Since SNe generally appear red
in $B-V$, especially after the peak, we recalculated the apparent
absolute magnitude limits in Table \ref{tab:photom} using a source
spectrum with $f_\nu \propto \lambda^{2.5}$. Synthetic SN light curves
were created from the $B$-band data of SN 1998bw published in
\citet{galama} and \citet{mkz99}. For a given peak brightness (allowed
to vary from $-15$ to $-35$\,mag) we randomly selected 1000 explosion
dates before the GRB and then constructed a set of synthetic light
curves, noting the frequency of synthetic precursors brighter than at
least one of the detection limits.  For simplicity we assumed no
correlation in peak brightness with light curve shape.

\begin{figure*}[hbt]
\centerline{\psfig{file=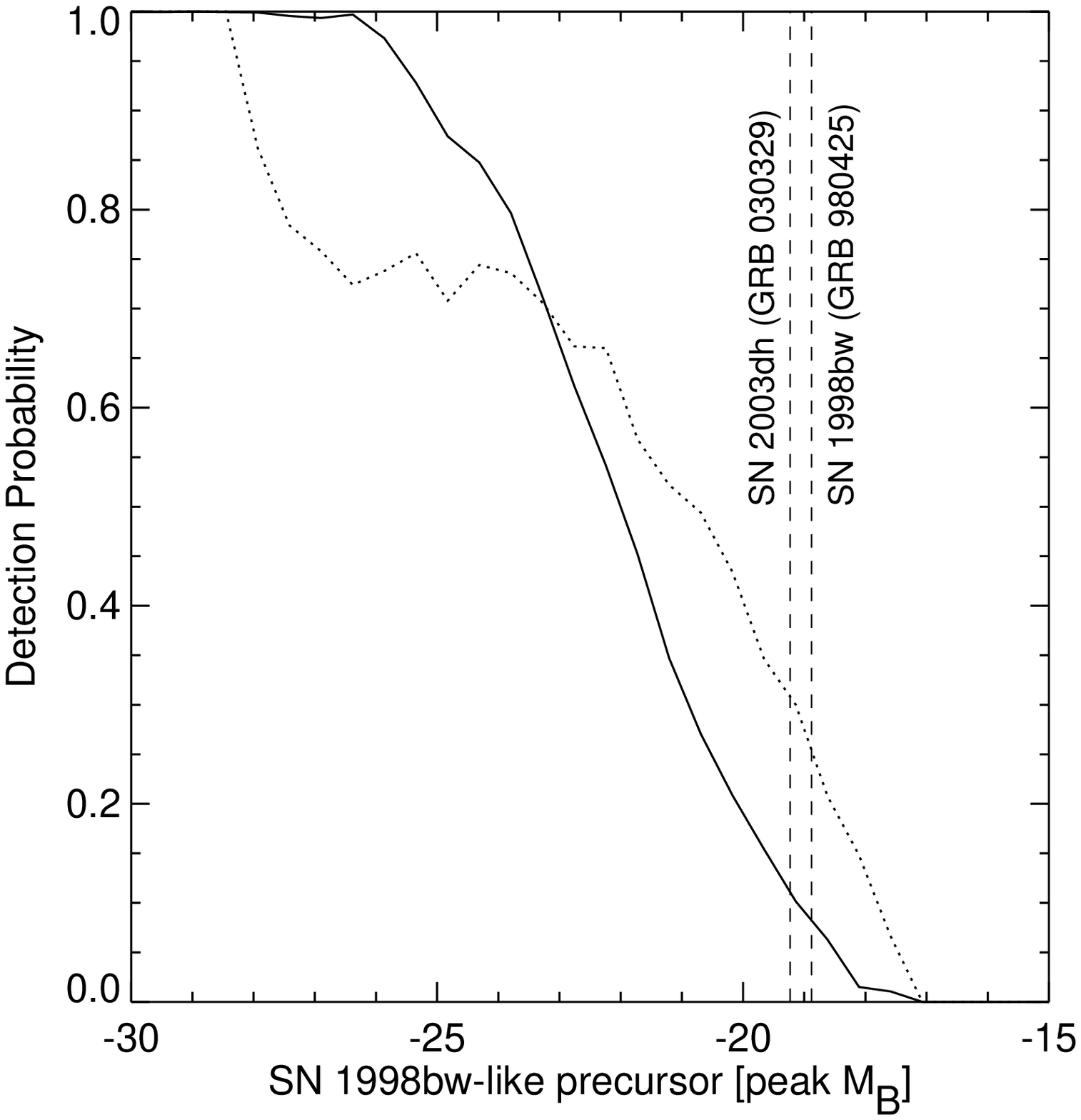,width=6.3in}}
\caption[]{Detection probability window for supernovae-like precursors
to GRBs. The solid curve shows the probability as function of peak
magnitude that an SN-like precursor with an explosion date randomly
distributed from 0 to 1000 days before a GRB would have been detected
in the NEAT data. The dotted curve shows the same probability except
that the explosion date was chosen uniformly in the logarithmic
interval from 1000 to 1 day before the GRB. Dashed vertical lines show
the peak magnitude of 1998bw (from \cite{mkz99}) and 2003dh, the SN
associated with GRB\,030329 (following from \cite{BLOOM03-ANDICAM-PAPER})}
\label{fig:sne} 
\end{figure*}

For two different distributions of starting times, we computed the
probability that the precursor would have been found in the NEAT data,
taking the ratio of the number of ``detected'' sources in the Monte
Carlo to the total number in the simulation. The result is seen in
Figure \ref{fig:sne}. For uniformly random explosion dates before the
GRB, almost all SN-like precursors brighter than (peak) $M_B \approx
-25$\,mag would have been detected at some point during the 3 years before 
the GRBs in the sample. For both preburst SN explosion date
distributions, the limits on SNe with 1998bw-like peak magnitudes
($M_B = -18.88$\,mag;
\citealt{mkz99}) is not very strong. Following Figure \ref{fig:sne}, there is only
about 10--30\% chance that such SNe would have been
detected. Therefore, though there are specific time windows when SNe
could not have occurred before the GRBs in the sample, we cannot rule
out the possibility that a SN with brightness comparable to 1998bw (or
2003dh) occurred {\it for all GRBs} but was missed due to the sparse
sampling of the data.

\subsubsection{Power-Law Precursors}

GRB precursors could, of course, be of a different physical nature
than supernovae. We have also tested the detection sensitivity to
decaying power-law precursors. We used a model for the flux of the
precursor as $f_\nu \propto (t - t_0)^\beta$ (for $t > t_0 + 1$ day), with 
the magnitude of the source at one day since $t_0$ of $M_B$(1 day). As 
above, we
simulated 1000 precursors with a characteristic start time $t_0$ less
than 1000 days before the GRB set. The spectral index of the precursor
was assumed to be $1$, for consistency with the conversion of apparent
magnitude to absolute $B$-band magnitude. For a range of decay
indices, $\beta$, and $M_B$(1 day), we found the probability of
detection using the results in Table \ref{tab:photom}.

\begin{figure*}[hbt]
\centerline{\psfig{file=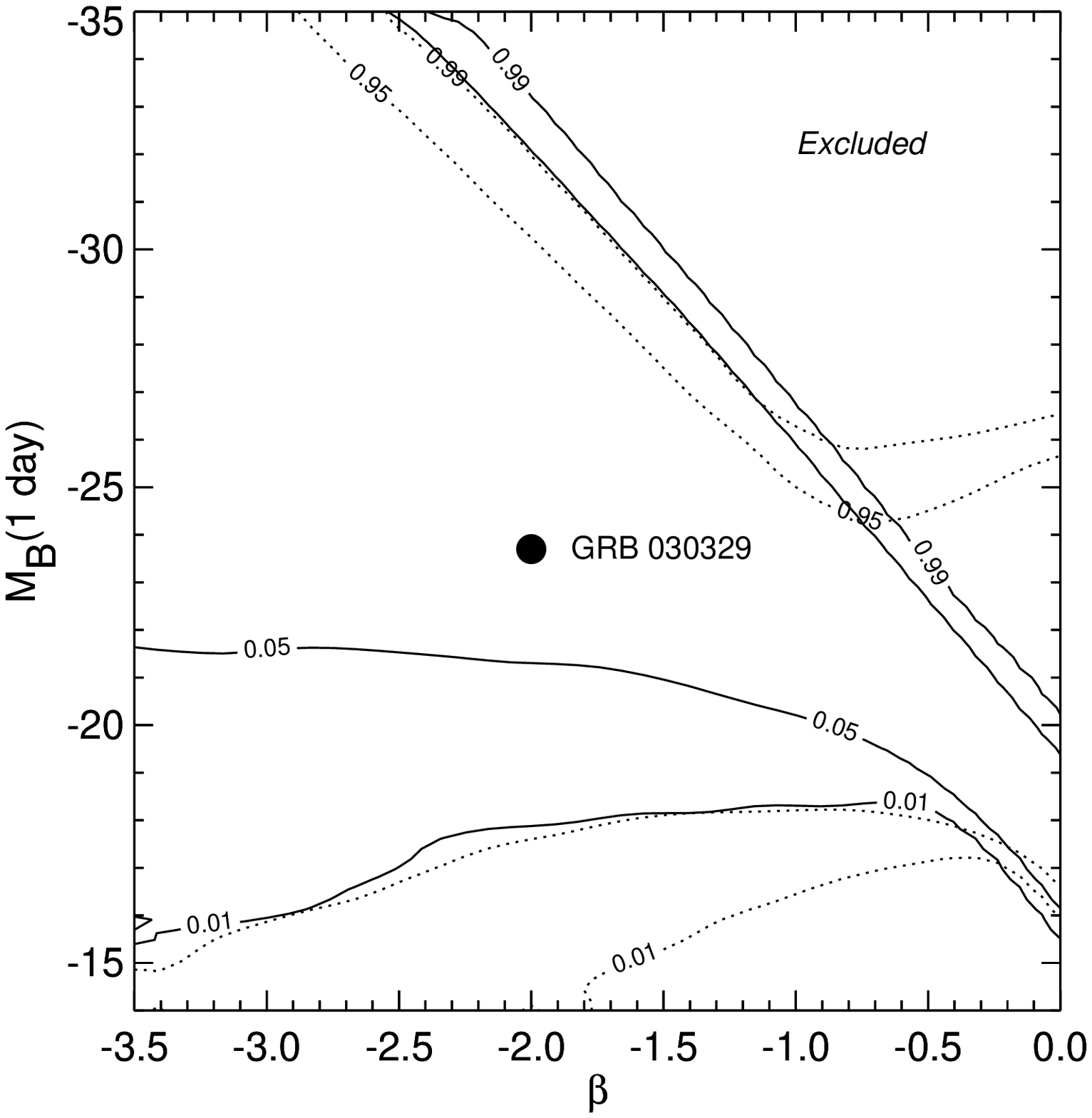,width=6.3in}}
\caption[]{Detection window for power-law precursors of GRBs. For each decay
index $\beta$ and $M_B$(1 day) we simulated random precursor start
times $t_0$ (solid lines: $t_0$ uniformly selected from 0 to 1000 days
before the GRBs; dotted lines: $t_0$ uniformly selected
logarithmically from 1 to 1000 days before the GRB). Shown are the
contours for 1\%, 5\%, 95\%, and 99\% detection
probabilities. Following from these results, if all GRBs had
precursors with $\beta \age -1$ and $M_B$(1 day) $\ale -25$ within
1000 days of the burst, then there is only a 1\% chance that the NEAT
data would have missed such an event. There is a $\sim$50\% chance of
missing precursors if all precursors behaved like the GRB 030329
afterglow.}
\label{fig:f4} 
\end{figure*}

Figure \ref{fig:f4} shows the results for two different probability
distributions of $t_0$. The lines of 0.95 and 0.99 show the equivalent
windows for detection of most such transients. For example, if all
precursors have $\beta = -1$ and $M_B$(1 day) $=-30$\,mag, then there
is greater than a 99\% chance that the NEAT set would have detected at
least one such precursor if it occurred less than 1000 days before the
GRB. Precursors that fade more quickly easily evade detection in the
sample. For example, only about 5\% of precursors with $M_B$(1 day)
$\approx -20$\,mag and $\beta < -1.5$ would be detected in the current
sample. Thus, there is little constraint on the existence of faint
precursors generated from tidal disruption around a massive black hole
($\beta = -5/3$; \citealt{lnm02}).

It is of interest to compare the brightness sensitivity of power-law
precursors to known afterglow brightnesses. Following from the
GRB\,030329 afterglow measurement of
\citet{GCN2056} and the extinction measurement in 
\citet{BLOOM03-ANDICAM-PAPER}, the restframe extinction corrected 
$B$-band magnitude of the afterglow was $M_B$(1 day) $\approx -23.7$,
fairly typical of other long-duration GRBs. Though there were
deviations from a power-law, the secular decay after 0.5 days was
approximately $\beta = -2$ \citep{pfk+03}. Following from Figure
\ref{fig:f4}, had a precursor -- perhaps of a similar physical origin as GRB
afterglows (a "foreglow") -- with the same parameters occurred for
all bursts at some time less than 1000 days before the GRB, there is
only a 50\% chance that such precursors would be seen in the NEAT
imaging.

\section{Discussion and Conclusions}

\begin{figure*}[hbt]
\centerline{\psfig{file=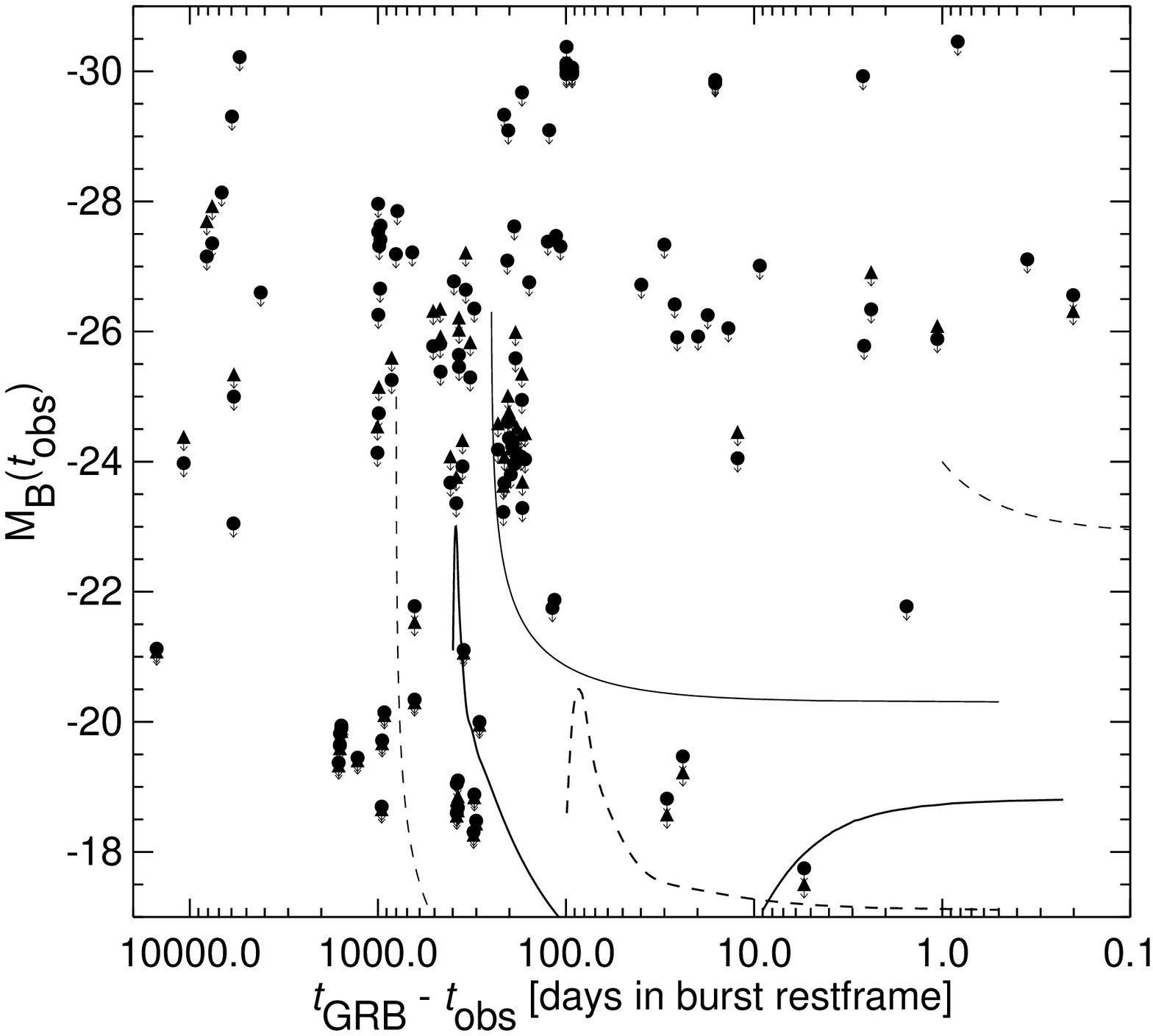,width=6.3in}}
\caption[]{Example SN-like and power-law precursor light curves near the
detection limit of the sample. The upper limits used in the Monte
Carlo test are shown for the power-law test (circles) and SN test
(triangles; $z_{\rm GRB} < 1.2$). The small differences are due to the
assumed spectrum of the precursor. Precursors shown are (from left to
right): power-law ($\beta = -1.3$, $M_B$(1 day) $=-25$\,mag, $t_0 =
-800$ day), SN-like ($M_B$(peak) $=-23.0$, $t_{\rm SN} = -400$ day),
power-law ($\beta = -1.0$, $M_B$(1 day) $=-26.3$\,mag, $t_0 = -250$
day), SN-like ($M_B$(peak) $=-20.5$, $t_{\rm SN} = -100$ day), SN-like
($M_B$(peak) $=-19.0$, $t_{\rm SN} = -10$ day), and power-law ($\beta
= -1.5$, $M_B$(1 day) $=-24$\,mag, $t_0 = -2$ day). Solid curves show
those precursors which would have been detected by at least one
photometric observation (and therefore cannot have occurred for all
GRBs in the sample). Dashed curves are for those precursors that could
have occurred for every GRB but would have escaped detection.}
\label{fig:f5} 
\end{figure*}

The non-detection of at least some power-law precursors---at
magnitudes comparable to afterglow brightnesses---supports the notion
that GRBs are due to explosive catastrophic events rather than due to
recurrent activity (e.g., from an AGN, \citealt{gyof+02}). For
consistency with all the limits presented herein, precursor
theories must now posit optical precursor emission fainter than $M_B
\approx -25$\,mag ($t_{\rm obs} - t_{\rm GRB} < 1$ day). 
From the analysis in \S 3, we can also rule out the possibility that
SN-like precursors with $M_B$(peak) $\ale -25$ and explosion dates
less than 1000 days after the bursts occurred for all GRBs in the
sample. Likewise, we exclude the possibility that all GRBs in the
sample had a precursor with $M_B$(1 day) $\ale -25$ and a slowly
fading ($\beta \ale -1$) power-law light curve. It is possible that a
subset of GRBs had precursors at least this bright. While providing
the strongest comprehensive precursor limits thus far, despite the
nearly 200 photometric pre-imaging measurements of 11 GRBs, the NEAT
data are not very sensitive to realistic precursor brightnesses. Even
if all GRBs had extremely bright [$M_B$(peak) $\approx -20$] SN-like
precursors, these would have been missed with an $\sim$80\%
probability. Similarly, more than half of precursors with similar
power-laws and brightnesses as GRB afterglows would have been
missed. Figure \ref{fig:f5} shows how various precursors could have
been detected or could have evaded detection.

Since GRB hosts are known to be forming stars vigorously, novae and
SNe should not be uncommon in such hosts. If the star-formation rate
in a GRB host is 40 $M_\odot$ yr$^{-1}$ then the canonical SN rate is
about 1 yr$^{-1}$ from somewhere within the host. At spatial
resolutions of $1-2$ arcseconds (such as from NEAT), the best
astrometric localizations of a precursor should be $\sim 200$
milliarcseconds ($1$-$\sigma$). This corresponds to a 3-$\sigma$
localization uncertainty 4\,kpc in projection at a redshift of $z =
0.5$, larger than the exponential scale lengths of all GRB hosts
measured to-date \citep{bkd02}. Therefore, any precursor emission
could be easily confused with an unrelated SN somewhere in the host
(this essentially extends the confusion problem encountered by plate
archive studies from the arcminute to the sub-arcsecond level). This
disturbing possibility motivates the additional need for high spatial
resolution in pre-imaging surveys. To this end, future deep wide-field
imaging surveys, such as PanSTARRS, should be sufficient: a precursor
detected with a signal--to--noise of 30 at the Nyquist-sample
resolution of PanSTARRS (0.6\arcsec) will result in a positional
confusion about 100 times smaller than with NEAT. This means that if
PanSTARRS detects an apparent precursor to a GRB within the host
galaxy (say one year before the burst), a significant association with
the GRB itself should be possible for even the most vigorously star forming
host galaxies.

The quality and quantity of ``pre-observations'' of transient
locations will dramatically improve with the increasing size of the
National Virtual Observatory\footnotemark\footnotetext{See {\tt
http://www.us-vo.org/}}. Digital imaging surveys, such as
RAPTOR \citep{vbb+03} and BOOTES \citep{ctsb+99}, have already presented
limits acquired a few hours before bursts to the $R \sim$ 12--15\,mag
level \citepeg{ctupm+03,wv03}. Deeper surveys such as SDSS should be
searched systematically for pre-imaging observations of GRB
locations. Identification of precursors with SDSS will have the added
benefit of color information, yielding important clues about the
physical nature of precursors, should any be detected.

The higher localization rates of {\it Swift} will increase the likelihood
of deeper pre-imaging with the next generation of large-area surveys
\citep{heyl03}. With the advent of large synoptic imaging programs (LSS and
PanSTARRS), with the primary science goals to detect near-earth
asteroids, the deepest and most frequent imaging will be at low
ecliptic latitudes. This observing strategy will be
well-matched to satellite missions like HETE-II \citep{rhl+02}
which preferentially localize GRBs anti-solar. Thus, the rates of deep
pre-imaging may be higher than those predicted by \citet{heyl03}. If
some sub-population of bursts originate from within the Galaxy or in
the local group (such as, potentially, short bursts and
supernovae-GRBs), then, as has been demonstrated with a handful of SNe
(see \citealt{vdlf03}), space-based pre-imaging might someday
retroactively resolve the progenitors directly.

\acknowledgements
We are exceedingly gratefully to the NEAT team (S.~Pravdo,
D.~R.~Rabinowitz, E.~F.~Helin, K.~Lawrence, T.~McGlynn, L.~Angelini,
and N.~White) for making images public and so readily
accessible. Without their dedication to this wonderful experiment and
user-friendly archive, none of this work would be possible. We thank
S.~Kulkarni for bringing to our attention a concern from P.~Kumar
about the possibility of misidentifying the origin of an SN bump {\it
after} a GRB, due to confusion with unrelated supernovae in the host
galaxies; our discussion of precursor confusion follows from that
concern.  We would also like to thank S.~Kulkarni, B.~Paczy\'nski, and
D.~Fox for helpful comments on the manuscript. J.S.B.\ is supported by
a Junior Fellowship to the Harvard Society of Fellows and by a
generous research grant from the Harvard-Smithsonian Center for
Astrophysics. C.B. would like to thank the Berkeley DEEP team for
their hospitality during the completion of this work. We thank our
physics teacher at Bethlehem Central High School, Mr.~Neff. We would also
like to thank an anonymous referee for their excellent suggestions for 
improvements to our original manuscript.


%
%
%

\newpage
\begin{deluxetable}{lccccc}
\tabletypesize{\footnotesize}
\tablewidth{4.3in}
\tablecaption{Data on 11 GRBs with Known Redshift and NEAT Precursor\label{tab:limits}}
\tablecolumns{6}
\tablehead{
\colhead{GRB} & \colhead{R-band limit\tablenotemark{a}} & \colhead{RA} & \colhead{DEC} & \colhead{$z$} & \colhead{$E(B-V)$}\\
\colhead{} & \colhead{mag} & \colhead{J2000} & \colhead{J2000} & \colhead{} & \colhead{mag} }
\startdata
030329 &  21.10 & 10:44:50.0  & +21:31:17    & 0.169 &  0.025\\
030323 &  21.31 & 11:06:09.4  & $-$21:46:13    & 3.372 &  0.049\\
030226 &  21.49 & 11:33:04.9  & +25:53:56    & 1.986 &  0.019\\
021211 &  22.01 & 08:08:59.9  & +06:43:38    & 0.800 &  0.027\\
021004 &  21.69 & 00:26:54.7  & +18:55:41    & 2.328 &  0.058\\
020813 &  21.59 & 19:46:41.9  & $-$19:36:05    & 1.254 &  0.110\\
020405 &  20.26 & 13:58:03.1  & $-$31:22:22    & 0.695 &  0.110\\
010222 &  19.85 & 14:52:12.6  & +43:01:06    & 1.477 &  0.023\\
000911 &  20.55 & 02:18:34.4  & +07:44:29    & 1.059 &  0.118\\
000418 &  20.62 & 12:25:19.3  & +20:06:11    & 1.118 &  0.033\\
000301C&  19.67 & 16:20:18.6  & +29:26:36    & 2.037 &  0.052\\
\enddata

\tablecomments{Redshift references for bursts up to and including GRB\,021004 are given in \citet{bloom2003}. The following references apply to the newer measurements: 
\citet{pbfl02} (GRB\,021211), \citet{pfd+03} (GRB\,030226), \citet{vwrh03} (GRB\,030323), and
\citet{gpe+03} (GRB\,030329).}

\tablenotetext{a}{The $2\sigma$ $R$-band upper limit at the position of the transient, derived from the sum of all NEAT pre-imaging data for the field.}
\end{deluxetable}

\newpage
\begin{deluxetable}{lcccccc}
\tabletypesize{\footnotesize}
\tablewidth{5.7in}
\tablecaption{Optical Pre-imaging Limits of 11 GRBs\tablenotemark{a}\label{tab:photom}}
\tablecolumns{7}
\tablehead{
\colhead{$t_{\rm obs}$\tablenotemark{b}} & \colhead{$t_{\rm GRB} - t_{\rm obs}$} & \colhead{Exposure} & \colhead{$R$-band\tablenotemark{c}} & \colhead{$B$-band\tablenotemark{c}} & \colhead{OT Flux} & \colhead{$\sigma$\tablenotemark{d}}\\
\colhead{UT Date} & \colhead{day} & \colhead{time (N x sec)} & \colhead{mag} & \colhead{mag} & \colhead{ADU} & \colhead{ADU}}
\startdata
 \\ \cutinhead{ GRB\,030329 ~~~$t_{GRB} =$ 2003-03-29 11:37:15
 ~~~$E(B-V)= 0.025$~~~~$z = 0.169$} 
2002-04-26 09:10:48 &   337.10 & 3x20 &   18.95 &   19.88 & 123.8 & 174.6\\
2002-04-12 09:41:59 &   351.08 & 3x20 &   20.47 &   20.82 & -11.5 & 41.2\\
2002-04-04 08:09:35 &   359.14 & 3x20 &   20.13 &   20.77 & 28.4  &57.0 \\
2002-04-01 07:00:00 &   362.19 & 3x60 &   20.64 &   20.53 & 20.5 & 111.3\\
2002-02-13 13:18:36 &   408.93 & 2x60 &   17.86  &   18.80 & 169.7 & 122.8\\
2002-01-14 10:52:48 &   439.03 & 3x20 &   20.28  &   20.93 & -9.2 & 51.1\\
2002-01-08 11:23:24 &   445.01 & 3x20 &   20.35 &   21.02 & -42.9 & 51.7\\
2000-04-13 14:15:00 &  1079.89 & 2x20 &   18.65 &   19.64 & -171.3 &  141.1\\
2000-04-13 06:37:48 &  1080.21 & 3x20 &   18.80  &   19.87 & -47.6 & 390.9\\
2000-03-14 14:15:00 &  1109.89 & 2x20 &   19.25  &   20.33 & -225.6 & 201.7\\
2000-03-08 11:09:35 &  1116.02 & 2x20 &   20.26  &   21.06 & -49.7 & 97.0\\
1999-02-19 12:30:00 &  1498.96 & 3x20 &   19.43  &   19.87 & 225.1 & 102.6\\
1998-03-27 09:30:00 &  1828.09 & 3x20 &   19.13  &   20.02 & 560.8 & 213.6\\
1998-03-26 09:23:24 &  1829.09 & 3x20 &   19.07 &   19.91 & -386.3 & 209.5\\
1998-02-23 10:23:24 &  1860.05 & 3x20 &   19.11 &   20.06 & 139.4 & 195.9\\
1998-02-22 10:23:59 &  1861.05 & 3x20 &   19.43 &   20.11 & -199.1 & 163.5\\
1998-01-24 12:50:24 &  1889.95 & 3x20 &   19.54 &   20.10 & 81.9 & 139.9\\
1955-03-25 06:00:00 & 17536.23 & 1x300 &  18.16 &   18.93 & 203.1 & 2037.3
\enddata

\tablenotetext{a}{The complete version of this table is in the electronic edition of the Journal. The printed edition contains only a sample.}

\tablenotetext{b}{Mean time of observation epoch.
 All observation dates before 1995 are from photographic imaging
 during the Palomar Sky Survey, scanned into digital form in the
 Digitized Sky Survey II. Observation dates after 1995 are from the
 NEAT experiment on the same telescope.}
\tablenotetext{c}{The 3-$\sigma$ upper limit to detection of precursor emission at the position of the GRB. These magnitudes have been calibrated using field stars and the USNO-A2.0 Catalog (see text). These limits do not include a correction for Galactic extinction (given as $E(B-V)$ from \citealt{sfd98}).}
\tablenotetext{d}{The $1~\sigma$ error on the measured flux at the OT position.}
\end{deluxetable}

\clearpage

\begin{figure*}[hbt]
\centerline{\psfig{file=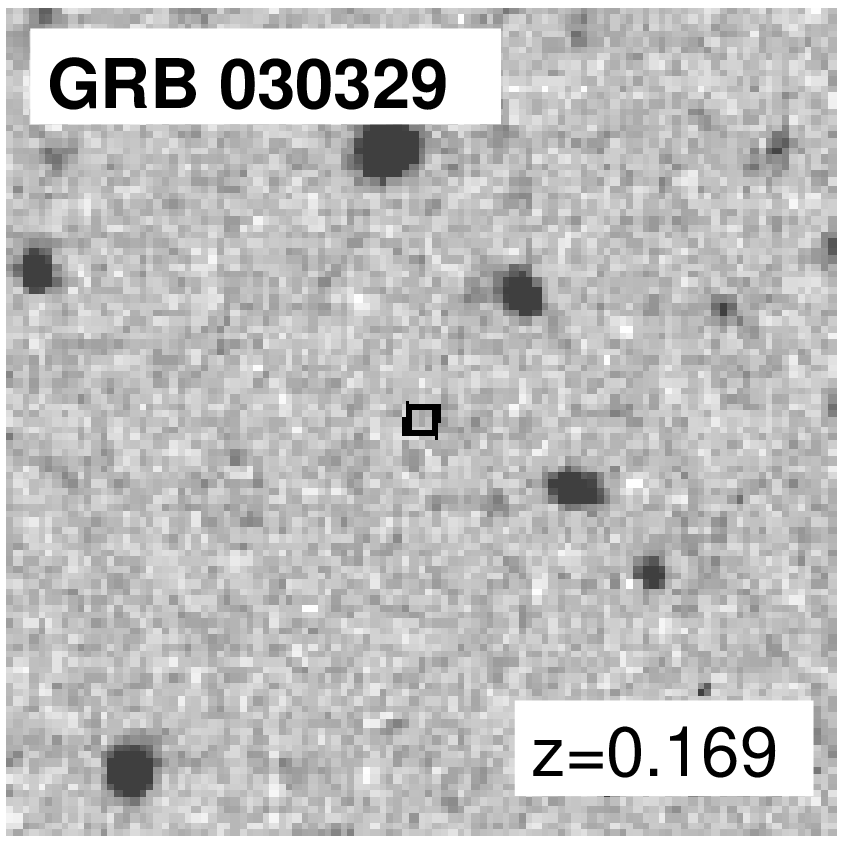,width=2.2in}
            \psfig{file=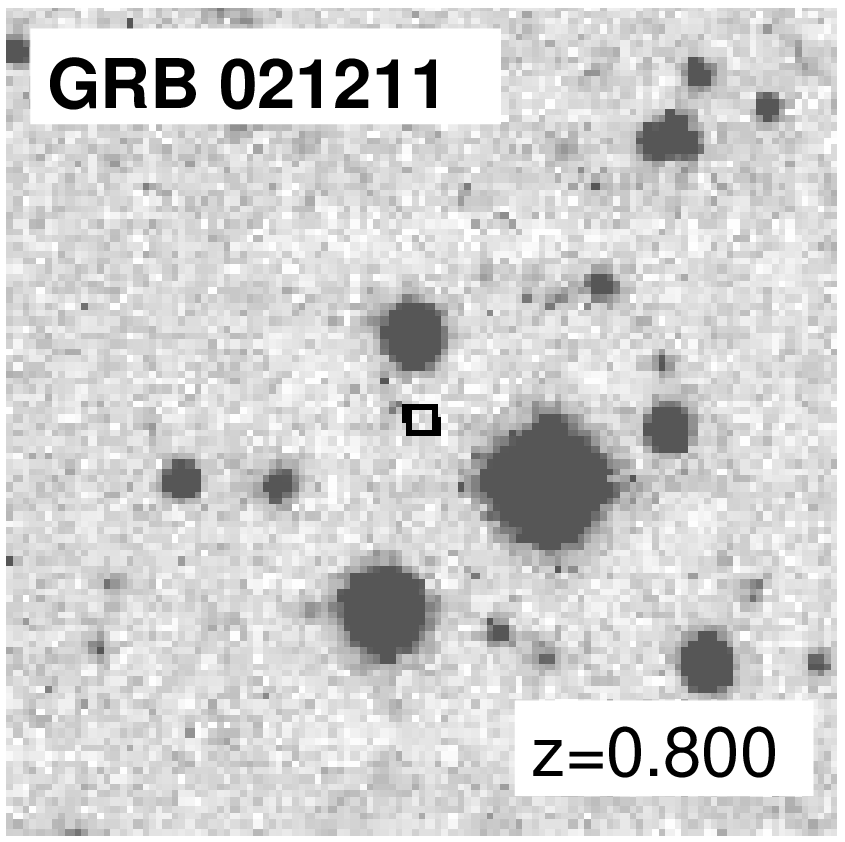,width=2.2in}}
\centerline{\psfig{file=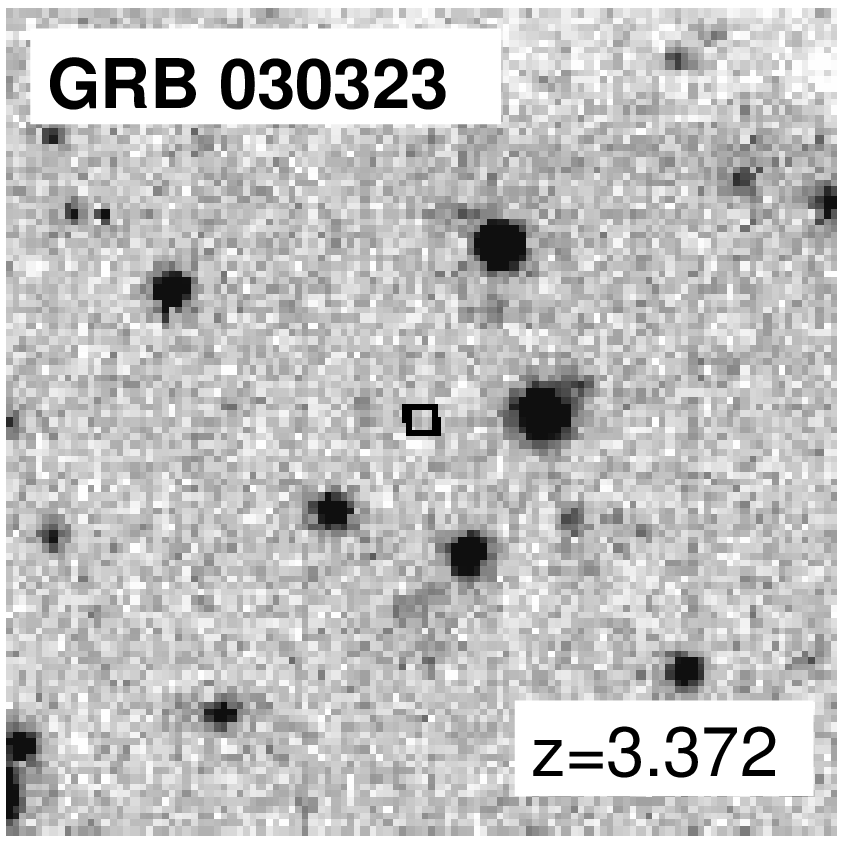,width=2.2in}
            \psfig{file=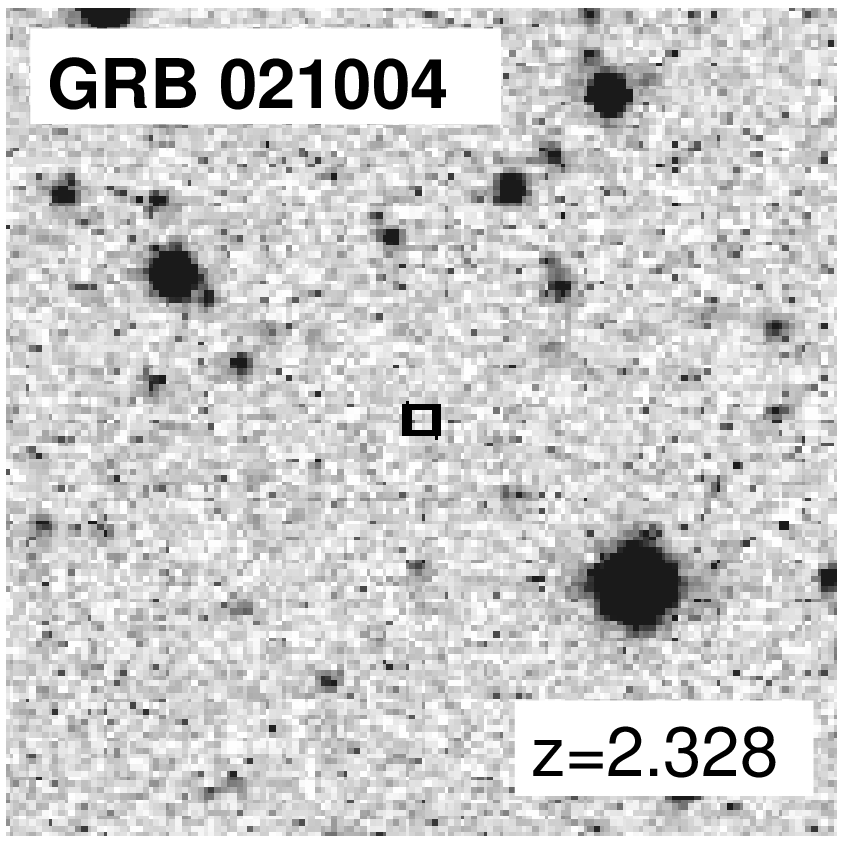,width=2.2in}}
\centerline{\psfig{file=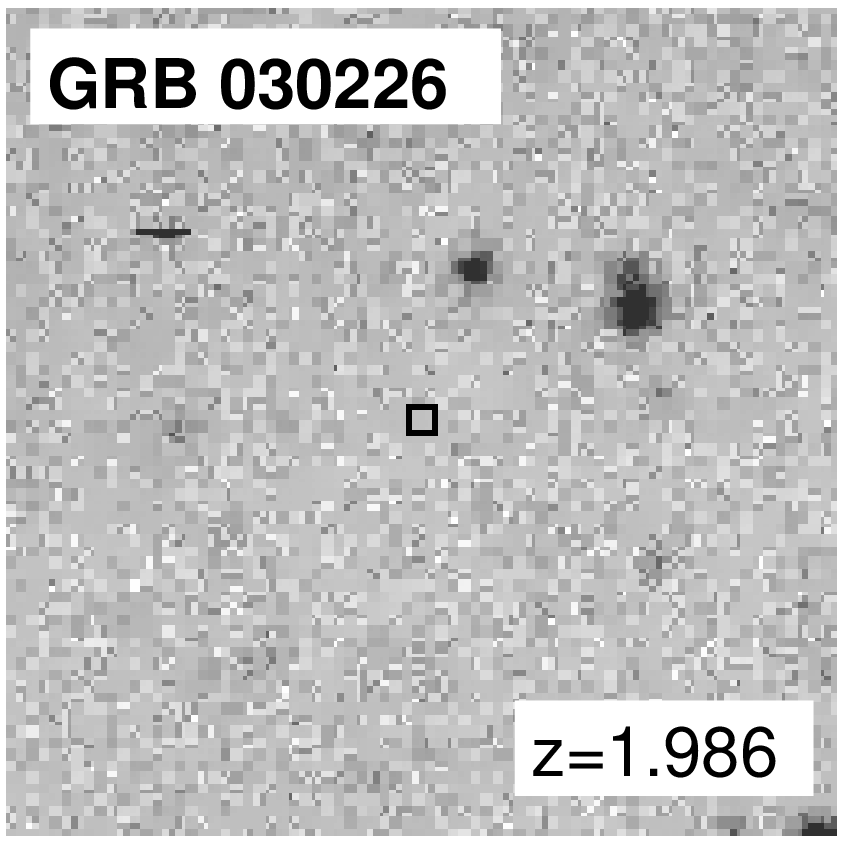,width=2.2in}
            \psfig{file=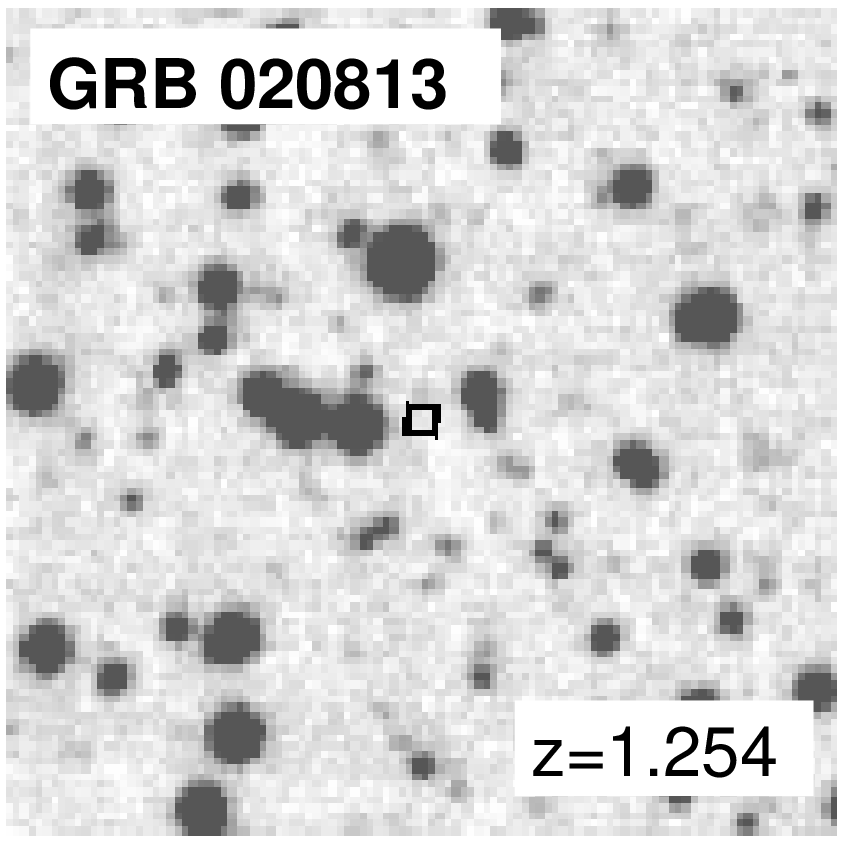,width=2.2in}}
\caption[]{Stacked preimages of the fields surrounding the 11 
GRBs with known redshift, presented achronologically from the trigger
date. Images are 150 $\times$ 150 arcsec$^{2}$ with North up and East
left. The 5$\times$5 arcsec box shows the position of the GRB afterglow
location.  The typical r.m.s.~astrometric error on our derived plate
solution is 0.5 arcsec in both axes.}
\label{fig:ims}
\end{figure*}

\newpage
\begin{figure*}[hbt]
\figurenum{\ref{fig:ims}}
\centerline{\psfig{file=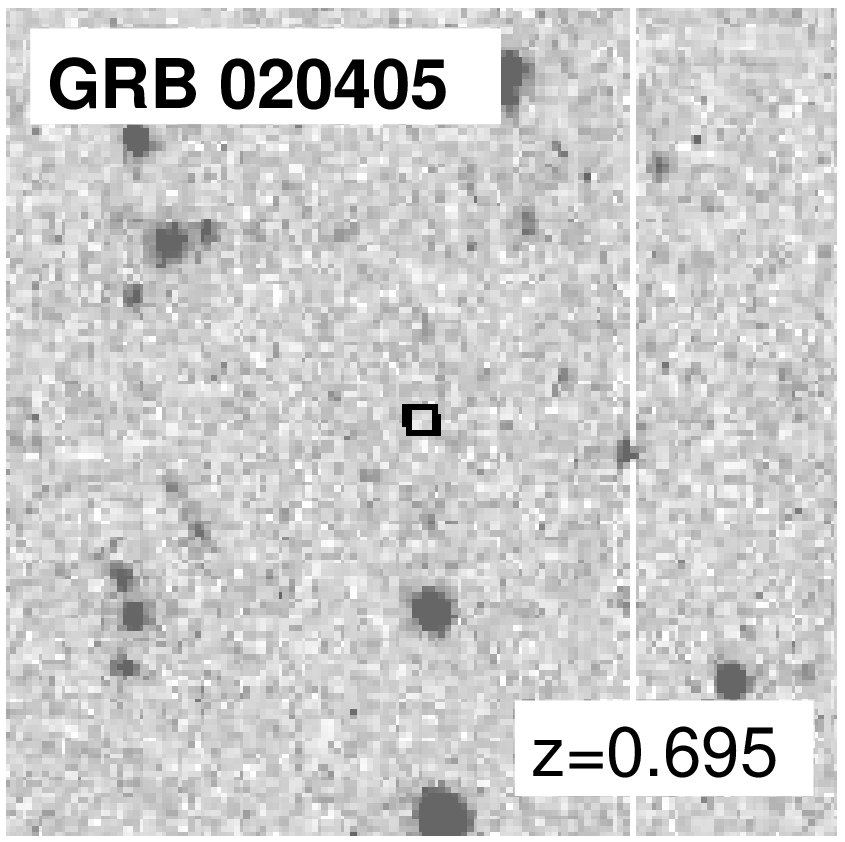,width=2.2in}
            \psfig{file=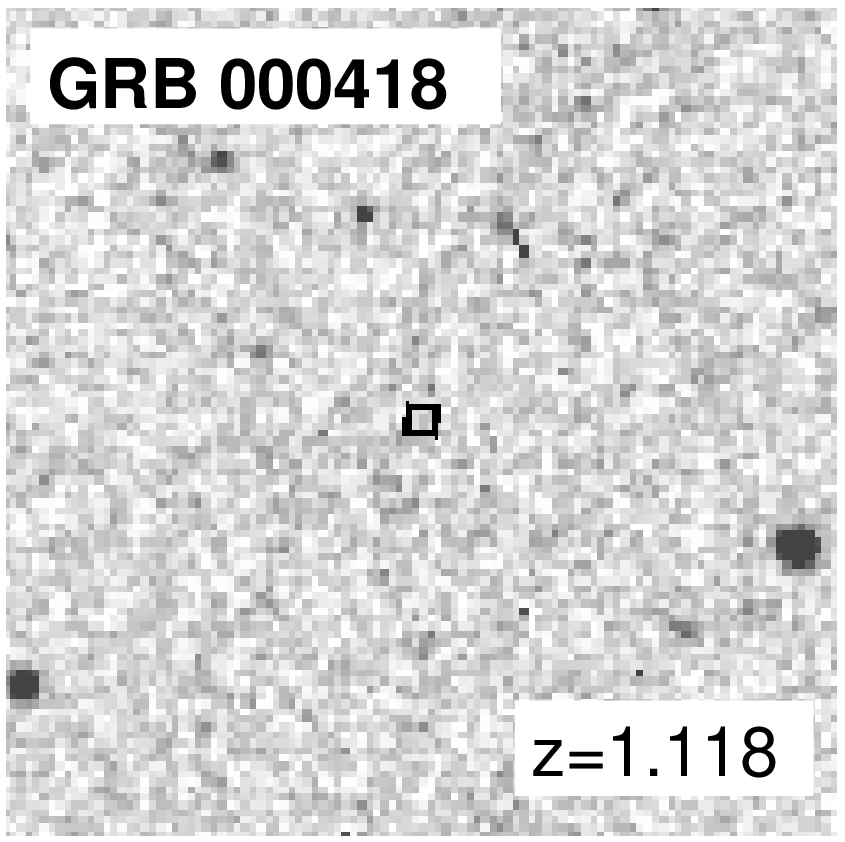,width=2.2in}}
\centerline{\psfig{file=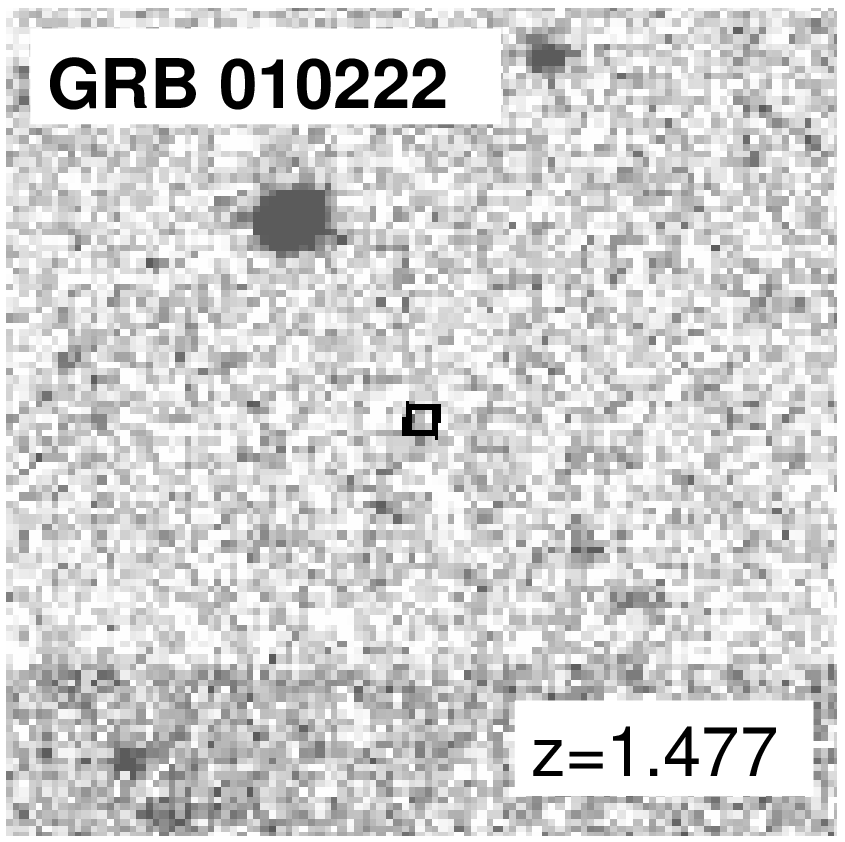,width=2.2in}
            \psfig{file=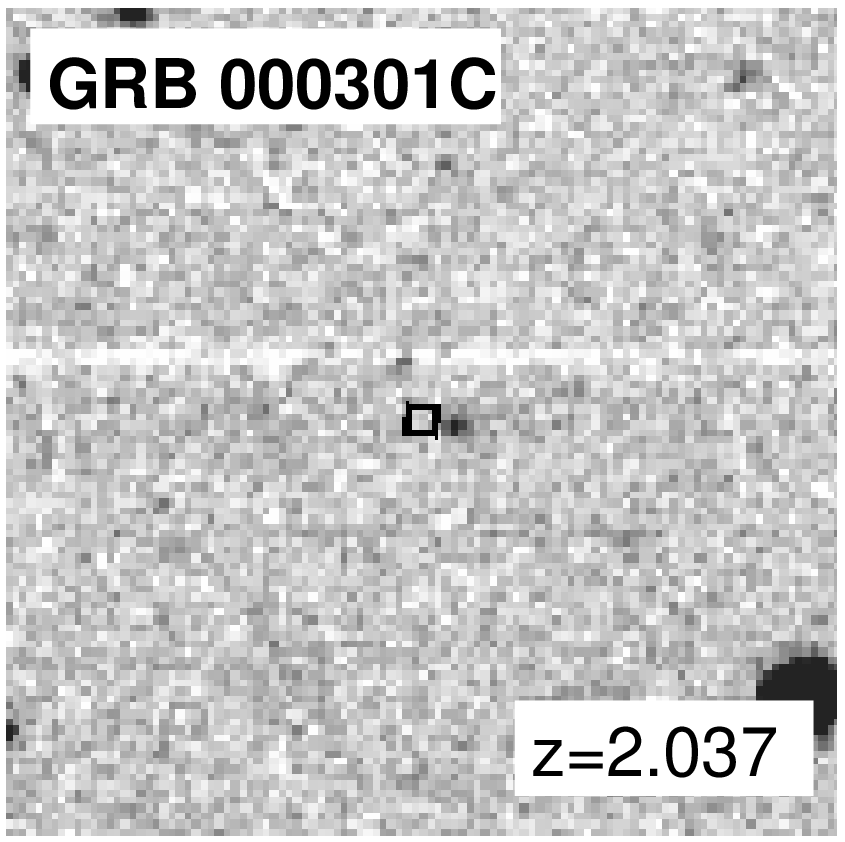,width=2.2in}}
\centerline{\psfig{file=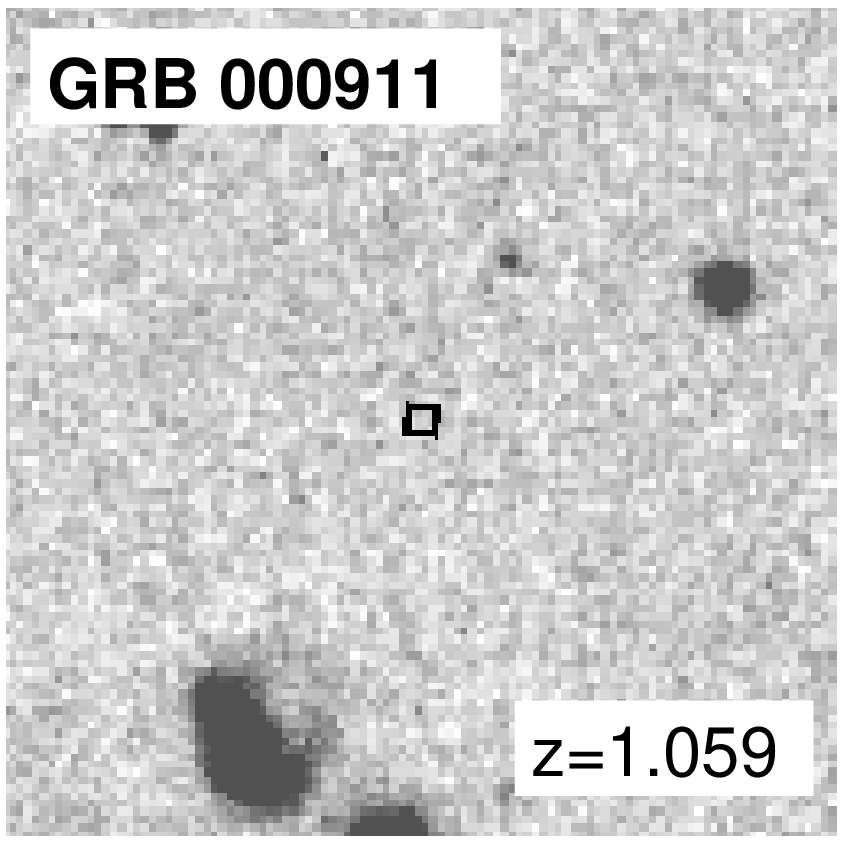,width=2.2in}}

\caption[]{(cont.) Stacked preimages of the fields surrounding the 11 
GRBs with known redshift, presented achronologically from the trigger
date. }
\label{fig:ims2}
\end{figure*}

\newpage

\end{document}